\documentclass[a4paper,11pt]{article}

\pdfoutput=1

\usepackage{jcappub}

\usepackage[english]{babel}
\usepackage{units}
\usepackage{siunitx}
\usepackage{graphicx}
\usepackage{amsmath}
\usepackage{amssymb}
\usepackage{subfigure}
\usepackage{lineno}
\usepackage{multirow}
\usepackage{tikz}
\usepackage{pgf}
\usepackage{xspace}

\definecolor{dark-gray}{gray}{0.3}

\title{Reconstructing the cosmic-ray energy from the radio signal measured in one single station}
\author[a,b,1]{C.~Welling\note{Corresponding author.}}
\author[c]{C.~Glaser}
\author[a,b]{A.~Nelles}

\affiliation[a]{DESY, Platanenalle 6, Zeuthen, Germany}
\affiliation[b]{Erlangen Center for Astroparticle Physics (ECAP), Friedrich-Alexander-Universit\"at Erlangen-N\"urnberg, Germany}
\affiliation[c]{University of California Irvine, USA}

\emailAdd{christoph.welling@desy.de, christian.glaser@uci.edu, anna.nelles@desy.de}

\abstract{Short radio pulses can be measured from showers of both high-energy cosmic rays and neutrinos. While commonly several antenna stations are needed to reconstruct the energy of an air shower, we describe a novel method that relies on the radio signal measured in one antenna station only. Exploiting a broad frequency bandwidth of $80-300$ MHz, we obtain a statistical energy resolution of better than 15\% on a realistic Monte Carlo set. This method is both a step towards energy reconstruction from the radio signal of neutrino induced showers, as well as a promising tool for cosmic-ray radio arrays. Especially for hybrid arrays where the air shower geometry is provided by an independent detector, this method provides a precise handle on the energy of the shower even with a sparse array. 
}

\begin{document}

\maketitle
\flushbottom

\section{Introduction and scientific motivation}
There is still no commonly accepted answer to the question of the origin of ultra-high energy cosmic rays. Detecting the radio emission of showers may bring us closer to an answer. Either through the detection of air showers, exploiting the excellent energy and composition resolution that comes with the radio method \cite{Aab:2016eeq,AERAPRD,Buitink:2016nkf,Bezyazeekov:2018yjw, Gottowik_2018}, or through the detection of the neutrino counterpart of ultra-high energy cosmic rays. Neutrinos can either be generated in interactions at the sources (\emph{astrophysical neutrinos}), allowing for multi-messenger detection of for example gamma-rays and neutrinos \cite{Halzen:2016gng,Ahlers:2018fkn,Murase:2016gly} or as \emph{cosmogenic neutrinos}  \cite{Greisen:1966jv,Zatsepin:1966jv,Beresinsky:1969qj}. These neutrinos are created through the interaction of ultra-high energy cosmic rays during propagation with the cosmic microwave or other photon backgrounds, making the neutrino flux sensitive to the composition of cosmic rays \cite{Kotera:2010yn,Ahlers:2010fw,vanVliet:2019nse}. All models predict a low flux of these neutrinos beyond \SI{e16}{eV} meaning that current detectors like IceCube are too small to detect a significant flux \cite{Aartsen:2018vtx} at the highest energies. Due to the long attenuation length of radio waves of $\mathcal{O}(\SI{1}{km})$ in ice \cite{ARIANNA_AbsorptionLength, Attenuation_Greenland, AttenuationSouthPole} allowing for sparse instrumentation, radio neutrino detectors are a promising alternative. 

\subsection{Method}
When reconstructing particle showers, be it neutrino or cosmic-ray induced, typically two aspects are relevant: the energy and the shower development, i.e. the shower maximum for air showers and the vertex position for neutrinos. Radio detection of air showers has reached maturity already, where the energy is typically derived from reconstructing the footprint of the shower on ground \cite{AERAPRD,Nelles:2014gma,Codalema,LOPES_energy}. To this end, the footprint is sampled by many antennas. The radio footprint is typically smaller than the particle footprint, since it is primarily governed by the width of the Cherenkov cone, which depends on the distance to the shower maximum and not the energy of the shower. Especially for events with small zenith angles, this results in measurable signals on only a radius of the order of \SI{100}{m} around the shower axis. In order to sample this footprint with at least three antennas, the radio arrays have to be relatively densely spaced. A much wider spacing would be needed to cost-effectively cover large areas, as it is the case for e.g.~the radio component of the proposed AugerPrime upgrade to the Pierre Auger Observatory \cite{AugerPrime}. At a spacing on the order of \SI{1}{km} the chances of being able to sample the full footprint of an air shower are small, unless targeting very horizontal showers ($> 60^{\circ}$) \cite{AERA_InclinedShowers}.
A similar challenge presents itself to proposed radio-based in-ice neutrino detectors (e.g. \cite{RNO,ARIA}). With a detector spacing of more than \SI{1}{km} it is unlikely for the radio signal from a neutrino interaction to be detected by more than one station. Since radio-neutrino detectors with surface antennas are also sensitive to air showers, detecting cosmic rays and reconstructing their energy will be a valuable tool for validating neutrino arrays, much like using muons in optical-Cherenkov neutrino experiments. Air showers show signal characteristics very similar to neutrino showers, making them both a background and a calibration tool.

In order to fully exploit this calibration signal, radio detection of air showers will need new reconstruction methods for neutrino arrays. In turn, sparse air-shower arrays may also benefit from these methods.  We will present a method that derives the energy of an air shower from the signal in a single detector station, exploiting the shape of the frequency spectrum. This method, while carrying merit in itself, can also be seen as step towards energy reconstruction for neutrinos. For the neutrino reconstruction additional complexities such as a much wider Cherenkov angle, frequency-dependent attenuation \cite{ARIANNA_AbsorptionLength, Attenuation_Greenland, AttenuationSouthPole}, and a strongly changing index of refraction leading to curved paths will have to be treated. However, we anticipate that using the shape of the frequency spectrum will also be essential for the energy reconstruction of neutrinos.

\subsection{Experimental context}
Since all large radio neutrino arrays are still in the proposal stage \cite{Barwick:2014pca,ARA,ARIA,RNO}, we have developed a method based on CoREAS Monte Carlo simulations \cite{Coreas}. However, in order to embed this method in realistic conditions, we have adopted a station design used in the pilot-stage array of ARIANNA \cite{ARIANNAtrigger}, which has already been used for air shower detection \cite{ARIANNA_cosmicRays}. A similar surface component employing log-periodic dipole antennas (LPDAs) is foreseen for all proposed in-ice neutrino arrays, which makes this a solid choice for air-shower reconstruction \cite{RNO, ARIA}. We note that the detailed modelling of a detector is not critical for the success of the method, as long as a wide frequency range ($> 100$ MHz) is detected.

For the purpose of this article, we adopt a station layout of four upward-facing LPDAs buried in the snow at a depth of roughly \SI{2}{m} below the surface. The antennas are distributed around a central tower, which in practice might be used for power and communications, but simply serves as symmetry center for this example. LPDAs on opposite sides of the central tower are \SI{8}{m} apart from each other and form a pair, as they have the same orientation and therefore measure the same polarization. One pair measures the north-south, the other the east-west polarization of the radio signal.

We adopt typical ARIANNA hardware characteristics. They include a very broad-band LPDA, which is filtered with a \SI{80}{MHz} high-pass and a \SI{500}{MHz} low-pass filter. After amplification the signal is digitized by an analog-digital converter (ADC) with a sampling rate of \SI{1}{GHz}. We also use the ARIANNA-style dual-threshold trigger scheme with a $4\sigma$ coincidence between at least two channels required for the station to trigger. As noise is an important factor for radio arrays, we use recorded noise from the set-up in Moore's Bay \cite{ARIANNA_cosmicRays}. Adopting ARIANNA characteristics allows us to study the influence of \emph{real} hardware characteristics as well as background effects to avoid an idealized study of the reconstruction capabilities.  In App.~\ref{sec:append_sites} we describe cross-checks for other locations on Earth, to exclude that the height and geomagnetic field at the ARIANNA site are a relevant factor.

In this analysis, we only use data from a single station and neglect possible additional information from other stations detecting the same air shower. While there will be some station coincidences in an array with a spacing in the order of \SI{1}{km}, especially for horizontal showers, we use this conservative estimate to develop the reconstruction algorithm. The reconstruction of an individual shower will of course improve, if information from two or more stations is provided. 

\section{Radio Signal Reconstruction}
The first step in the event reconstruction is the retrieval of the electric field from the voltage traces measured in the antennas, which in turn have to be retrieved from the recorded ADC counts. To do so, we use the NuRadioReco software framework. A  detailed description of NuRadioReco and its event reconstruction methods is presented elsewhere \cite{NuRadioReco}, so only the reconstruction of the signal direction and the electric field will be covered here briefly, as they are essential for the energy reconstruction.
\subsection{Arrival direction}
The distance of \SI{8}{m} between two antennas with the same orientation is small compared to the size of the air shower footprint at ground level. Therefore the radio signal measured in the two channels of an antenna pair can be expected to be very similar, only shifted by the difference in signal travel time between the antennas.

We define the correlation
\begin{equation}
    \label{eq:correlation}
    \rho (\Delta n) = \frac{\sum_i (V_{1})_i\cdot (V_{2})_{i-\Delta n}}{\sqrt{\sum_i(V_{1})_i^2}\cdot\sqrt{\sum_i(V_{2})_i^2}}
\end{equation}
between the voltage time traces $V_1$ and $V_2$ of the two channels in each pair, shifted by $\Delta n$ samples relative to each other. The correlation $\rho(\Delta n)$ is normalized to $-1<\rho(\Delta n)<1$ to make it independent of the signal amplitudes.

Assuming a plane-wave front, we calculate the expected time difference and the corresponding $\Delta n$ between antennas of each channel pair for a given signal direction and calculate $\rho(\Delta n)$. The direction of the recorded signal is determined by finding the direction for which the sum of $\rho(\Delta n)$ of both channel pairs is largest. Using this method, an angular resolution better than \SI{1}{^\circ} is achieved for signals that exceed a signal-to-noise ratio of 4\footnote{We define the signal-to-noise ratio (SNR) as the ratio between half the peak-to-peak amplitude and the noise RMS.}.

\subsection{Electric field}

\begin{figure}
    \centering
    \includegraphics[width=\textwidth]{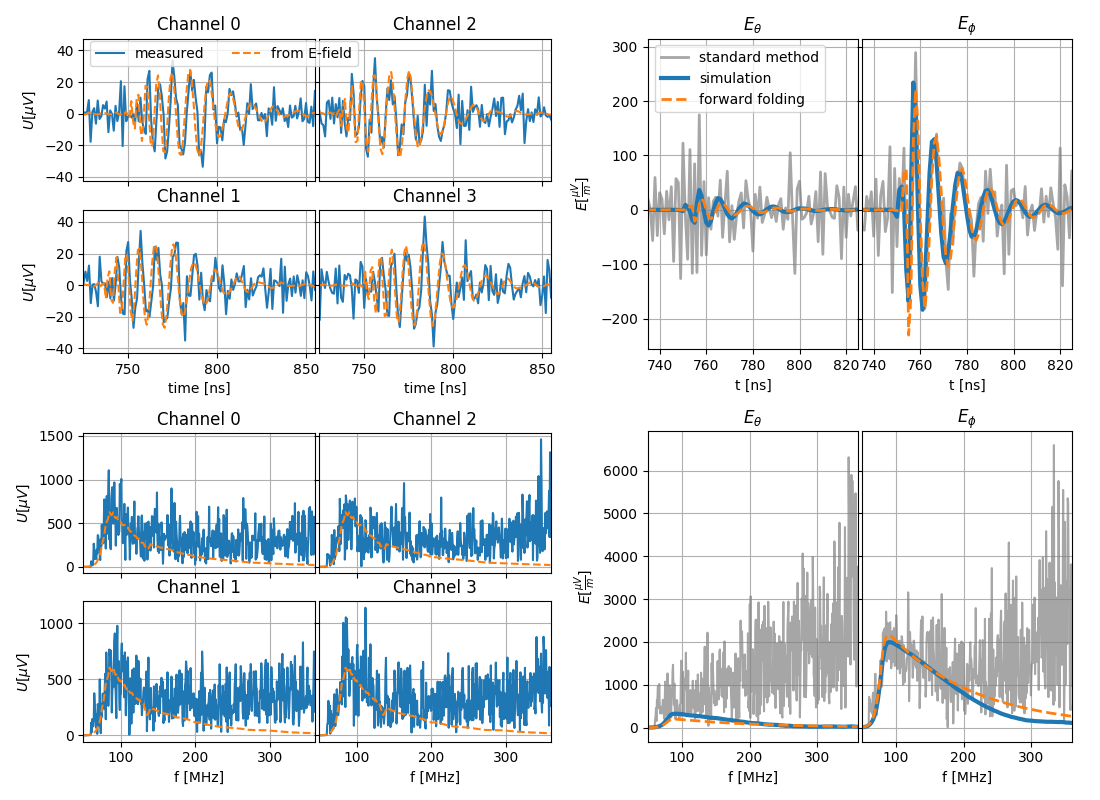}
    \caption{Example of the electric-field reconstruction using the standard and the forward folding technique. Left panels: Voltage traces of four spatially displaced antennas. Shown are both the time- (top) and frequency domain (bottom). The solid blue curve represents the measured voltages whereas the dashed orange curve shows the analytic solution of the forward folding technique. Channels 0,2 and 1,3 are parallel, the measured signal only differs in noise contribution.  Upper right panels: Reconstructed electric field trace using both the standard (gray) and forward folding (dashed orange) technique in comparison to the simulated true (solid blue) values. Lower right panels: Reconstructed amplitude spectrum using both techniques in comparison with the simulated truth (color scheme same as above).}
    \label{fig:forward_folding_example}
\end{figure}

With the incoming direction of the radio signal known, it is possible to reconstruct the 3-dimensional electric field trace from the voltage traces measured by the antennas. Since the distances between a station's antennas are small, we can assume that all four channels measured the same electric field. Then, in Fourier space, the voltage traces in the antennas for a given electric field $\mathcal{E}^{\theta, \phi}$ are given by
\begin{equation}
    \begin{pmatrix} \mathcal{V}_1(f) \\ \mathcal{V}_2(f) \\ ...\\ \mathcal{V}_n(f)\end{pmatrix} = 
    \begin{pmatrix} \mathcal{H}_1^\theta (f)& \mathcal{H}_1^\phi (f)\\ \mathcal{H}_2^\theta (f) & \mathcal{H}_2^\phi (f)\\ ... \\ \mathcal{H}_n^\theta (f)& \mathcal{H}_n^\phi (f)\end{pmatrix} 
    \begin{pmatrix} \mathcal{E}^\theta(f) \\ \mathcal{E}^\phi(f)\end{pmatrix} \ + \text{noise} ,
    \label{eq:H_full}
\end{equation}
where $\mathcal{H}_i^\theta$ and $\mathcal{H}_i^\phi$ are the antenna responses of the i\textsuperscript{th} antenna to the $\theta$ and $\phi$ components of the electric field. Eq.~\eqref{eq:H_full} is an overdetermined system of linear equations that can be solved for $\mathcal{E}^\theta(f)$ and $\mathcal{E}^\phi(f)$ via a $\chi^2$ optimization to reconstruct the electric field. However, this method does not respond well to noise in the voltage traces \cite{NuRadioReco}. We therefore use the so-called forward folding method instead, which provides a more accurate electric field reconstruction, especially with regards to the frequency spectrum by using an analytic model of the radio signal.

In the frequency domain, the electric field can be approximated by the equation
\label{sec:forward_folding}
\begin{equation}
     \begin{pmatrix} \mathcal{E}_\theta \\ \mathcal{E}_\phi \end{pmatrix} = \begin{pmatrix} A_\theta \\ A_\phi \end{pmatrix} 10^{f \cdot  m_f} \, \exp(\Delta\, j) \, 
\label{eq:pulse}
\end{equation}
that depends on only four parameters: The signal amplitudes $A_\theta$, $A_\phi$ of the two orthogonal polarizations, the frequency slope $m_f$ and a frequency-independent phase offset $\Delta$. $f$ is the frequency and $j$ the imaginary unit.

The electric field is reconstructed by calculating $\mathcal{V}_i(f)$ for a given $\mathcal{E}_\theta(f)$, $\mathcal{E}_\phi(f)$ using Eq.~\eqref{eq:H_full} and optimizing $A_\theta$, $A_\phi$, $m_f$ and $\Delta$ through a $\chi^2$ fit to the recorded voltage time traces. 
An example of such a forward folding reconstruction is displayed in Fig.~\ref{fig:forward_folding_example}. The voltage traces were generated by folding the electric field from an air shower simulated using CoREAS \cite{Coreas} with the detector response and adding a noise trace recorded by an ARIANNA station in the field. It shows that the forward folding method avoids the problem of overestimating of the electric field at higher frequencies that the standard method suffers from, resulting in a more accurate spectrum reconstruction. Especially the reconstruction of the frequency slope $m_f$ is much more accurate than with the standard method. For a detailed comparison between the performance of the two methods, see \cite{NuRadioReco}.

From the electric field trace we calculate the energy in the radio signal per unit area, called the energy fluence \cite{AERAPRD}
\begin{equation}
    \Phi_E=\epsilon_0\cdot c\cdot\Delta t\cdot\sum_i |\vec{E}(t_i)|^2
\end{equation}
where $\epsilon_0$ is the vacuum permittivity, $c$ is the vacuum speed of light and $\Delta t$ is the sampling rate. We note that we do not need to subtract noise from the energy fluence, because the forward-folding technique reconstructs an electric field free of noise. 

\section{Energy estimator}
Our goal is to use the energy fluence $\Phi_E$ as an estimator in order to measure the energy in the electromagnetic shower. In our scenario, each air shower is only detected by a single radio station and no additional information about the shower geometry is available. Thus, an energy reconstruction from the energy fluence at a single point is challenging, as it is influenced by a number of factors other than the shower energy: 
\begin{description}
\item [Angle to the magnetic field:]
The dominant emission process for air showers is the so-called geomagnetic emission, whose energy fluence is proportional to $\sin^2(\alpha)$ where $\alpha$ is the angle between the shower axis and  the geomagnetic field.
\item [Distance to the shower maximum:]
The radio signal gets weaker with increasing distance between the shower and the radio antenna. Since the largest part of the radio signal is emitted near the shower maximum, the signal energy fluence roughly scales with the geometric distance to the shower maximum $d_{X_{max}}$ squared.
\item [Viewing Angle:]
We define the viewing angle $\varphi$ as the angle between the shower axis and a line from the shower maximum to the antenna. Emission is strongest if $\varphi$ is close to the Cherenkov angle (defined by $\cos(\varphi_{Cherenkov})=\frac{1}{n}$, where $n$ is the index of refraction at the shower maximum) and becomes weaker the larger the difference between the two gets.
\end{description}
For an estimator to achieve a good energy resolution, we need to take these factors into account and correct for them.

\subsection{Procedure}
Correcting for the angle to the magnetic field is rather straightforward, since the shower axis is very close to the signal direction and the magnetic field at any site can be measured directly. For the correction, it is useful to introduce a coordinate system that has one axis aligned with the shower axis $\vec{v}$, one in the direction of the Lorentz force $\vec{v}\times\vec{B}$ and the third one perpendicular to those two $\vec{v}\times(\vec{v}\times\vec{B})$. The energy fluence can then be split into the energy fluence of the two polarizations $\Phi_{\vec{v}\times\vec{B}}$ and $\Phi_{\vec{v}\times(\vec{v}\times\vec{B})}$. Since only the geomagnetic emission, which is polarized in the $\vec{v}\times\vec{B}$ direction, is affected by the angle to the magnetic field, only $\Phi_{\vec{v}\times\vec{B}}$ is corrected for this effect.

The distance $d_{X_{max}}$ to the shower maximum is harder to account for, because the atmospheric depth $X_{max}$ of the shower maximum is unknown. Fortunately, however, this effect is most relevant for inclined showers, where the shower maximum is far away from the observer. In this case, event-by-event fluctuations of $X_{max}$ have little effect and $d_{X_{max}}$ is mainly a function of the zenith angle \cite{GeoCELDF}. It is therefore sufficient to assume a typical value for $X_{max}$ and calculate the expected $d_{X_{max}}$ based on the shower inclination. Based on \cite{AverageXmax} we use an average $X_{max}$ of \SI{750}{\frac{g}{cm^2}}\footnote{In fact, the value chosen for $X_{max}$ has very little effect on the energy reconstruction. If a different $X_{max}$ had been chosen, the zenith-dependent parameterization of the effect of the viewing angle (Sec.~\ref{Sec:Ccone}) would compensate this.}. To make sure that this assumptions was correct, we checked for correlations between the uncertainties on the reconstructed energy and $X_{max}$ and found no such biases.

This lets us correct for those two effects by defining a corrected energy fluence
\begin{equation}
    \Phi_E^\prime = \bigg(\frac{1}{\sin^2(\alpha)} \Phi_{E,\vec{v}\times \vec{B}}+\Phi_{E,\vec{v}\times (\vec{v}\times \vec{B})}\bigg)\cdot \bigg(\frac{d_{X_{max}}}{\SI{1}{km}}\bigg)^2
\end{equation}
which will be used as the basis for the energy estimator. We note that we ignore the influence of the second order charge-excess emission process on the $\Phi_{E,\vec{v}\times \vec{B}}$-component as it is small enough to not limit the resulting energy resolution. Since geomagnetic and charge excess emission can interfere both constructively and destructively depending on where the observer is on the $\vec{v}\times\vec{B}$ axis, the effect of the charge excess emission will largely average out and not lead to a systematic offset. In the future, the term $\sin^2(\alpha)$ may be replaced by $a^2 + (1-a^2)\sin^2(\alpha)$ following the same approach as in \cite{Glaser:2016qso} where $a$ is the relative strength of the charge-excess component that depends on the distance to $X_{max}$ and the shower axis, and can be parameterized as function of zenith angle using the work of \cite{GeoCELDF}. 

The third correction for the effect of the viewing angle is more complex and has to be assessed in a Monte Carlo study. We use a first set of simulated air showers to derive the correction factor for the viewing angle and a second set to test the full energy reconstruction (see Sec.~\ref{sec:energy_resolution}). 

The Monte Carlo data set was produced using 545 CoREAS \cite{Coreas} simulations of proton-induced air showers and their radio emission. The showers cover an energy range of \SI{e16} - \SI{5e19}{eV} and random arrival with zenith angles between \SI{0}{^\circ} and \SI{80}{^\circ} and a uniform azimuth distribution. 
The energy range is set by the typical threshold of radio signals above the thermal and Galactic noise floor. This depends in detail on the experimental set-up but is typically at several PeV.

For each shower, the radio signal was simulated at 160 positions arranged in a star-like pattern around the shower core at an altitude of \SI{30}{m} above sea level (the height of the ARIANNA detector on the Ross Ice shelf), and the detector response to the radio signal at each position was simulated. Radio simulations are in general time consuming, but every observer position can be used independently providing additional statistics. Also, since the radio signal stems from the electromagnetic component of the air shower, which is in general rather smooth, small air shower sets provide enough information. We find that we are not statistics limited (see also App.~\ref{sec:append_sites}).

A simple-threshold trigger was simulated by requiring that at least two channels recorded a pulse larger than \SI{60}{mV} ($\sim 3 V_{rms}$) for the reconstruction to continue. A 10th order Butterworth bandpass filter with a passband of \SI{80}-\SI{300}{MHz} was applied and a full event reconstruction was performed, using the forward folding technique described in Sec.~\ref{sec:forward_folding}. The resulting data set was used to develop an energy estimator that can also account for the effect of the viewing angle.

\subsection{Correction for the viewing angle}
\label{Sec:Ccone}
\begin{figure}
\centering
\includegraphics[width=\textwidth]{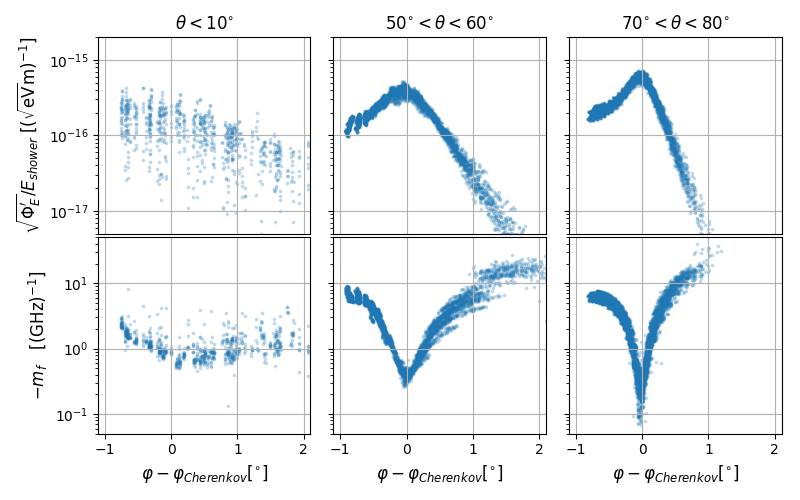}
\caption{Top row: Energy fluence of the radio signal over electromagnetic shower energy as a function of the viewing angle relative to the Cherenkov cone. Bottom row: Frequency slope parameter $m_f$ as a function of the viewing angle relative to the Cherenkov cone. Each column represents a different range of zenith angles $\theta$. The Cherenkov angle is calculated using the index of refraction of the air at the shower maximum.}
\label{fig:signal_ldf}
\end{figure}

The shape and amplitude of a radio pulse emitted by an air shower differs depending on the viewing angle $\varphi$ under which it is seen. Under small viewing angles, signals from different stages in the shower's development arrive simultaneously, leading to an amplified emission over a wide frequency band \cite{Werner2012}. The resulting dependency of the frequency spectrum on $\varphi$ has already been observed in the $30-80$ MHz  band \cite{JansenPhD,GrebePhD,LOFAR_spectrum, AERA_InclinedShowers} and is used to reconstruct air shower energies by the ANITA experiment \cite{ANITA_Energy}. The method used by ANITA, however, requires dedicated Monte Carlo simulations for each reconstructed event and is therefore not suitable for large arrays with regular detections.

The radio signal is strongest if the viewing angle coincides with the Cherenkov angle at the location of the shower maximum. In this case, the radio signals emitted at the shower maximum add up coherently, leading to an amplification of the signal. If the event is seen from further off the Cherenkov cone, overlapping signals lose coherence and the amplitude decreases, as can be seen in the top row of Fig.~\ref{fig:signal_ldf}. However, shorter wavelengths lose coherence faster than longer ones. The effect this has on the frequency spectrum can be observed with the frequency slope parameter $m_f$ (see Eq.~\ref{eq:pulse}), shown in the bottom row of Fig.~\ref{fig:signal_ldf}. The small absolute value of $m_f$ at the Cherenkov angle corresponds to a relatively flat frequency spectrum, while the larger $m_f$ off the Cherenkov cone corresponds to a sharp dropoff towards higher frequencies in the spectrum. The event-by-event fluctuations in $\sqrt{\Phi_E^\prime}/E_{shower}$ and $m_f$ are large for $\theta<10^\circ$, but they decrease with increasing zenith angle. It is also worth noting that neither the change in $\sqrt{\Phi_E^\prime}$ nor in $m_f$ is symmetric around the Cherenkov angle.

\begin{figure}
    \centering
    \includegraphics[width=\textwidth]{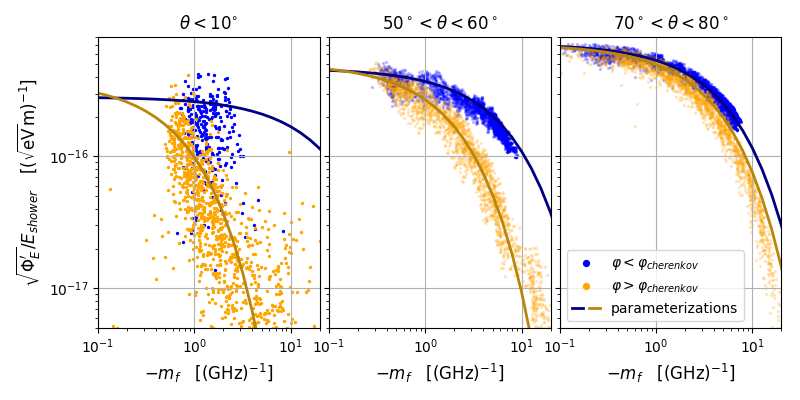}
    \caption{Energy fluence of the radio signal over electromagnetic shower energy as a function of the frequency slope parameter $m_f$ for different zenith angles $\theta$. Events are split into two categories: Those viewed from inside the Cherenkov cone are displayed in blue, those from outside in orange. The lines show the parametrization Eq.~\ref{eq:energy_fluence_parametrization} for the center of the zenith angle bin.}
    \label{fig:dinosaur_plot}
\end{figure}

Since $m_f$ is a function of $\varphi-\varphi_{Cherenkov}$, it is possible to use it as a proxy for the viewing angle. This is done in Fig. \ref{fig:dinosaur_plot}, which shows $\sqrt{\Phi_E^\prime}/E_{shower}$ as a function of $m_f$, with small $|m_f|$ corresponding to events seen by observers close to the Cherenkov ring and larger $|m_f|$ to those seen from further off the Cherenkov angle. Events are split into two groups: Those where the station is located inside the Cherenkov ring and those seen from outside of it. While $\sqrt{\Phi_E^\prime}/E_{shower}$ and $m_f$ show a clear correlation for larger zenith angles, the scatter for the $\theta<10^\circ$ events is very large. This is likely due to the small angle between the shower axis and the geomagnetic field, which causes the charge-excess emission to become more prevalent compared to the geomagnetic emission. Since both signal components have a different polarization, they interfere constructively at positions in positive $\vec{v}\times\vec{B}$ direction and destructively in negative $\vec{v}\times\vec{B}$ directions. This causes the lateral distribution function (LDF) of the radio signal to no longer be rotationally symmetric in the shower plane. Together with the smaller dependence on the viewing angle (Fig. \ref{fig:signal_ldf}), this leads to the large scatter (see also discussion in App.~\ref{sec:append_sites}).

\begin{figure}
    \centering
    \includegraphics[width=\textwidth]{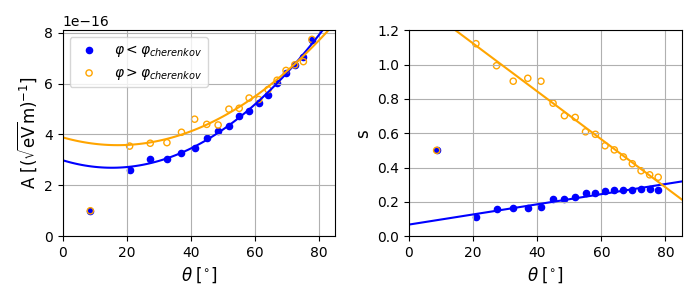}
    \caption{Results of the fits of Eq.~(\ref{eq:energy_fluence_parametrization}) for the different zenith bins and. A quadratic and a linear function were fitted to the zenith dependence of the parameters $A$ and $s$, respectively, and are shown as solid lines. Results for events seen from inside the Cherenkov ring are shown in blue, those seen from outside in red. Zenith angle bins $\theta<30^\circ$ were excluded from the fits.}
    \label{fig:fit_parameters}
\end{figure}

We divide the events into 20 zenith angle bins with equal sky coverage and fit the function
\begin{equation}
    \frac{\sqrt{\Phi_E^\prime}}{E_{shower}}= A \cdot \exp(-s\cdot (|m_f|\cdot\mathrm{GHz})^{0.8})
    \label{eq:energy_fluence_parametrization}
\end{equation}
to the distribution in each bin separately for events seen from within and from outside of the Cherenkov cone. The results are shown in Fig.~\ref{fig:fit_parameters}. The zenith dependence of $A$ can be parameterized by a second-order polynomial, the one of $s$ by a straight line. The specific values for this parametrization are shown in Tab.~\ref{tab:parametrizations}.

Using these parameterizations, Eq.~\eqref{eq:energy_fluence_parametrization} can now be solved for $E_{shower}$ to obtain an energy estimator using the zenith angle $\theta$ and the radio energy fluence $\Phi_E$ and frequency slope $m_f$. However, since events viewed from inside the Cherenkov cone have to be treated separately from those seen from outside of it, a way to distinguish between the two is needed, which will be addressed in the next section.

\subsection{Inside vs.~outside of the Cherenkov cone}
\begin{figure}
    \centering
    \includegraphics[width=.5\textwidth]{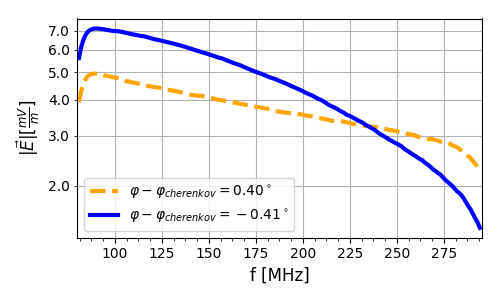}
    \caption{Spectra of the radio signals from the same air shower measured inside (blue) and outside (orange) of the Cherenkov ring. Both spectra deviate from the exponential parametrization used by the forward folding method, but while the spectrum curves upwards outside the Cherenkov ring, it curves downward inside of it. The sharp drop at \SI{80}{MHz} and \SI{300}{MHz} is caused by the bandpass filter.}
    \label{fig:spectrum_second_order}
\end{figure}
\begin{figure}
    \centering
    \includegraphics[width=\textwidth]{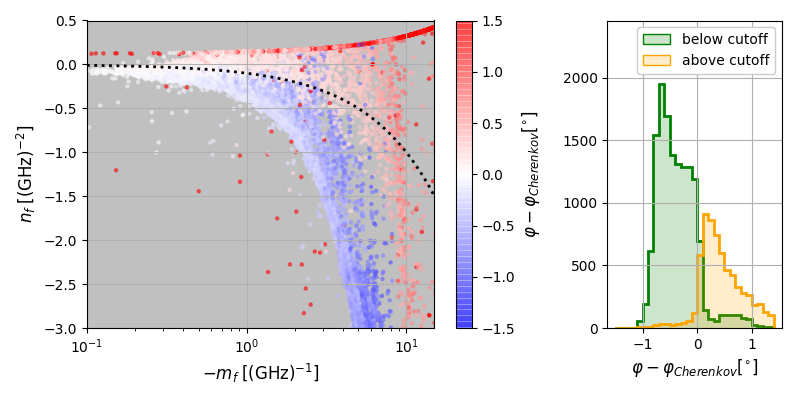}
    \caption{Left: Relation between the spectrum slope parameter $m_f$ and the quadratic correction parameter $n_f$ for events with zenith angles $\theta \leq 30^\circ$. The color signifies the viewing angle $\varphi$ relative to the Cherenkov cone, with events inside the Cherenkov cone in blue and outside in red. Right: Distribution of the viewing angles $\varphi$ relative to the Cherenkov cone for events above and below the black dotted cutoff line in the left plot.}
    \label{fig:second_order_correction}
\end{figure}
When reconstructing the electric field in Sec.~\ref{sec:forward_folding}, we approximated frequency dependence of $\log_{10}(|\vec{\mathcal{E}}(f)|)$ by a straight line. As is shown in Fig.~\ref{fig:spectrum_second_order}, the actual radio signal deviates from this, with the frequency spectrum curving downward at higher frequencies for events seen from inside the Cherenkov ring and upward for those seen from outside of it. While the difference is small enough that it does not need to be taken into account when reconstructing $\Phi_E$ and $m_f$, it can be used to distinguish between the two event categories.

We extend the reconstruction of the electric field by an additional step, in which we modify Eq.~\eqref{eq:pulse} by introducing an additional 2nd order term in the exponent:
\begin{equation}
     \begin{pmatrix} \mathcal{E}_\theta \\ \mathcal{E}_\phi \end{pmatrix} = \begin{pmatrix} A_\theta \\ A_\phi \end{pmatrix} 10^{f \cdot  m_f + (f-80\mathrm{MHz})^2 \cdot n_f} \, \exp(\Delta\, j) \, 
\label{eq:pulse_second_order}
\end{equation}
This pulse is fitted to the recorded voltage traces as described in Sec.~\ref{sec:forward_folding}, using the results from fitting Eq.~\eqref{eq:pulse} as starting values. The amplitudes $A_\theta$ and $A_\phi$ are allowed to change, while $m_f$ is kept constant.

The relation between $\varphi$, $m_f$ and $n_f$ is shown in Fig.~\ref{fig:second_order_correction}. Showers seen from within the Cherenkov ring tend to have a smaller $n_f$ than those seen from outside of it. A simple linear function (black dotted line) can be used to separate the events into two groups: Those above the line and those below it. Several cuts have been tested leading to similar efficiency and purity, so we have chosen to use the simplest. The histogram on the right shows the viewing angles relative to the Cherenkov cone of the two groups, demonstrating that these provide a good separation between events viewed from within and outside of the Cherenkov ring. A fraction of 90\% of the events are either identified correctly or are less than \SI{0.05}{^\circ} away from the Cherenkov angle. A mistake in the separation of events this close to the Cherenkov ring will result only in minor uncertainties on the reconstructed energy, as the energy parametrizations for both cases are very similar near the Cherenkov angle. One may notice that for large $|m_f|$, the discriminator seems to work very poorly, as events outside the Cherenkov ring tend to have small values of $n_f$. In principle, this could be improved by also defining a maximum value for $|m_f|$ above which all events are classified as being outside the Cherenkov ring. In practice however, the signals from these events are so weak that they are unlikely to be detected, making this additional step unnecessary.

As it was the case for the energy parametrization, this method works best for inclined showers. The difference in $n_f$ for showers seen from within and outside of the Cherenkov ring tends to increase with the zenith angle. For small zenith angles, both event categories mostly overlap.

Knowing if the station is inside or outside of the Cherenkov ring, Eq.~(\ref{eq:energy_fluence_parametrization}) can now be used to determine the electromagnetic shower energy. The resulting energy reconstruction algorithm has been implemented as the \emph{cosmicRayEnergyReconstructor} module in the \emph{NuRadioReco} software framework \cite{NuRadioReco}, where one can see the full implementation in detail and use it to reconstruct the energy accordingly \cite{cosmicRayEnergyReconstructor}.

\section{Obtainable energy resolution}
\label{sec:energy_resolution}
Having developed an energy parametrization and a way to distinguish between air showers seen from inside and outside the Cherenkov ring, these can now be combined to reconstruct the shower energy. By performing this energy reconstruction on simulated air showers and comparing the result to the true energy, we can determine its performance under realistic conditions.
\subsection{Monte Carlo data-set}
\begin{figure}
    \centering
    \includegraphics[width=\textwidth]{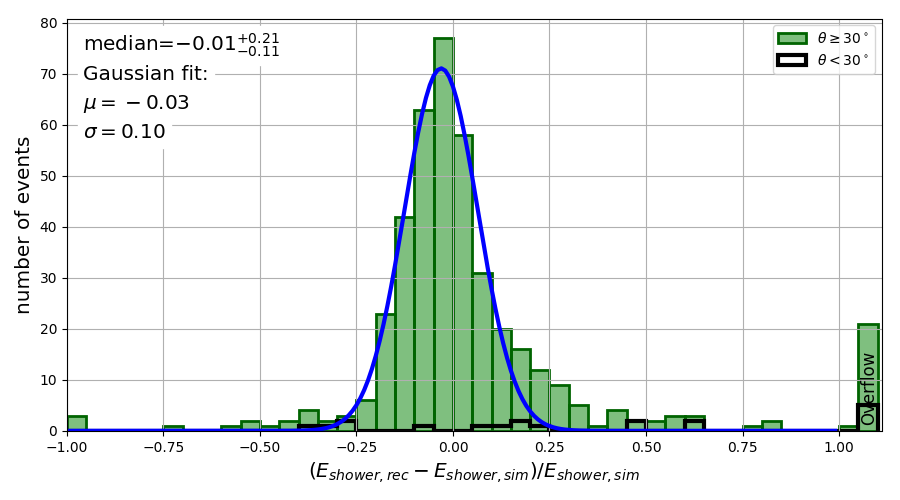}
    \caption{Uncertainties on the reconstructed energies for air showers with a zenith angle of $\theta\geq30^\circ$ (green) and $\theta<30^\circ$ (black). For $\theta\geq30^\circ$ the distribution has a median of $-0.01^{+0.21}_{-0.11}$ with the uncertainties representing the 68\% quantile. Fitting a Gaussian function to the histogram yields an average of $\mu=-0.03$ and a standard deviation of $\sigma=0.10$. Note that the rightmost bin is an overflow bin.}
    \label{fig:reconstruction_quality}
\end{figure}
To estimate the uncertainty of the reconstructed energies, a set of 200 CoREAS \cite{Coreas} simulations of air showers and their radio emissions was used. To mimic a realistic data set, the showers had an isotropic distribution of incoming directions up to a zenith angle of $80^\circ$ and energies in the range of $\SI{e17}{eV}<E<\SI{e20}{eV}$ with a spectral index of $-2$. For each air shower, the radio signal was simulated at 160 positions, arranged in a star-like pattern around the shower core. To achieve a realistic distribution of station positions, for each simulated air shower, 20 random positions within a radius of \SI{2}{km} around the shower core were produced. For each of these random positions, the simulated electric field closest to it was determined and used in the analysis. More detailed information on the used Monte Carlo set is provided in App.~\ref{sec:append_MCset}.

Simulated electric field traces were folded with the antenna response and the effect of the signal chain  was simulated to obtain the voltage traces. To account for the effect of noise, one of 100,000 forced-trigger events recorded by ARIANNA stations were randomly selected and its recorded voltage trace added. A dual-threshold trigger \cite{ARIANNAtrigger} was simulated by requiring the voltage pass both an upper and a lower threshold of $\pm\SI{80}{mV}$ (about 4 times the typical noise RMS) within \SI{5}{ns} in at least 2 channels. Of the 4,000 radio pulses, 597 passed this selection.
A 10th order Butterworth filter with a pass-band of \SI{80}-\SI{300}{MHz} was applied to the voltages and a reconstruction of the incoming direction and the electric field was performed. The electromagnetic shower energy was estimated from Eq.~\eqref{eq:energy_fluence_parametrization}. Since the electric field reconstruction corrects for the detector response to the radio signal and the same detector description was used for simulation and reconstruction, the effect of uncertainties on the hardware components is neglected here. These concerns are addresses separately in Sec. \ref{sec:hardware_uncertainties}.

\subsection{Energy resolution}
The resulting energy resolution is shown in Fig.~\ref{fig:reconstruction_quality}. While the energy reconstruction does not work well for vertical showers, for more inclined showers the distribution has a median of $-0.01^{+0.21}_{-0.11}$ with the uncertainties representing the 68\% quantile. While the uncertainty is small for most events, there are several outliers where the energy was greatly overestimated.
\begin{figure}
    \centering
    \includegraphics[width=\textwidth]{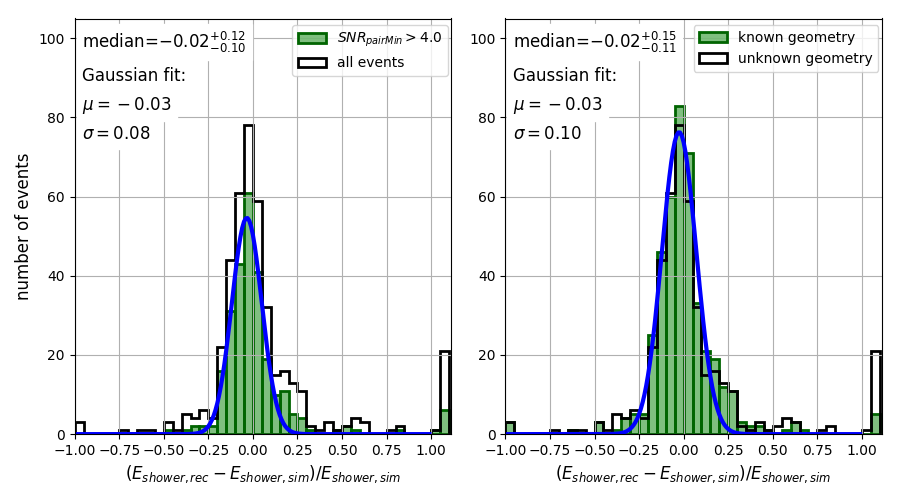}
    \caption{Left: Energy resolution for air showers with a zenith angle of $\theta\geq30^\circ$ if a cut requiring the signal-to-noise ratio of at least one channel in each pair to be larger than 4 is applied, compared to the resolution without any cuts. In this case, the median is $-0.02^{+0.12}_{-0.10}$ with the uncertainties representing the 68\% quantile. Fitting a Gaussian function to the histogram yields an average of $\mu=-0.03$ and a standard deviation of $\sigma=0.08$. Right: Achievable energy resolution if reliable information if the air shower was seen from inside or outside of the Cherenkov ring was available from another source. In this case, the distribution has a median of $-0.02^{+0.15}_{-0.11}$ with the uncertainties representing the 68\% quantile. Fitting a Gaussian function to the histogram yields an average of $\mu=-0.03$ and a standard deviation of $\sigma=0.10$.}
    \label{fig:alternative_resolutions}
\end{figure}
Most of these outliers happen with events that have a low signal-to-noise ratio. Because some periods have a higher noise level than normal \cite{AriannaHRA3DesignPerformance}, even a trigger with a threshold of \SI{80}{mV} can produce events with a small signal-to-noise ration. Another problem can occur with the direction reconstruction. Because each channel pair measures a different polarization, it is possible for the radio pulse to only be detected in one pair. In this case, the direction reconstruction is likely to produce a wrong result which in turn leads to a wrong reconstruction of the electric field and the energy. We therefore impose a stricter cut requiring that at least one channel in each pair has as $\mathrm{SNR}>4$. The resulting energy resolution is shown in Fig. \ref{fig:alternative_resolutions} (left). With this cut, most of the outliers disappear and the resolution improves, with a median of $-0.02^{+0.12}_{-0.10}$.

It turns out that most of the outliers in Fig.~\ref{fig:reconstruction_quality} are caused by events viewed from within the Cherenkov ring being mistaken for events seen from outside of it. While an ARIANNA-like detector has no other way to identify if the station was inside or outside the Cherenkov ring, this task may be trivial for other radio detector designs, for example those that are part of a hybrid detector. Fig.~\ref{fig:alternative_resolutions} shows the energy resolution if reliable information if the shower was seen from inside the Cherenkov ring or outside of it was available from some other source. In this case most outliers disappear and the resulting distribution has a median of $-0.01^{+0.15}_{-0.11}$ with the uncertainties representing the 68\% quantile, having applied no SNR-cut. The remaining outliers are still mostly events with a relatively low signal-to-noise ratio, in which the electric field reconstruction did not converge properly.

It should be noted that this method, like all radio methods, reconstructs the energy in the electromagnetic cascade of the shower \cite{Glaser:2016qso}. To get the cosmic ray energy, the invisible energy, which can be up to $\sim20\%$ of the total shower energy, has to be taken into account as well \cite{InvisibleEnergy}.

\subsection{Systematic effects of hardware uncertainties}
\label{sec:hardware_uncertainties}
\begin{figure}
    \centering
    \includegraphics[width=\textwidth]{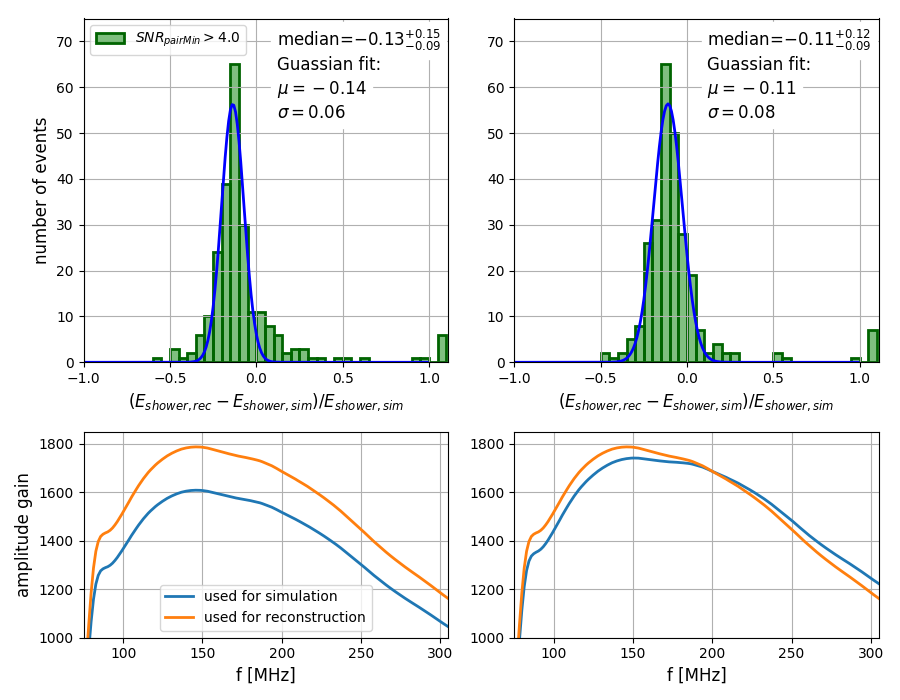}
    \caption{Effect of uncertainties in the amplifier response on the energy reconstruction. Top Left: Resolution of the reconstructed electromagnetic shower energy if the amplifier gain is scaled by a factor of $p=0.9$ (Eq.~\ref{eq:amp_scale}). Top Right: Resolution of the reconstructed electromagnetic shower energy if the amplifier gain is modified by Eq.~\ref{eq:amp_slope} with $q=0.1$. Bottom: Gain of the amplifiers used for the event simulation and reconstructions that resulted in the shower energy reconstructions shown above.}
    \label{fig:hardware_errors}
\end{figure}

Up to this point, the event simulation and reconstruction have used the same detector description, implying perfect knowledge of the detector response. This is, of course, not realistic, so in this section we will investigate the effect of hardware uncertainties on the energy resolution.

The procedure is as follows: We repeat the same simulation and reconstruction as in Sec.~\ref{sec:energy_resolution}, but when the detector response is simulated, certain aspects of the detector description are modified. Then, we reconstruct the event using the original detector description. This mimics the effect that uncertainties in the hardware would have on the event reconstruction.

The first investigated uncertainty is a deviation in the orientation of the antennas. We produce two sets of simulated events with one of the antennas rotated by $2^\circ$ and $5^\circ$ around the z-axis and perform a full event reconstruction using the default antenna orientation on both sets. The uncertainties on the reconstructed energies show no significant difference to those in Fig.~\ref{fig:reconstruction_quality}. We therefore conclude that the influence of uncertainties on the antenna orientation on the reconstructed energy is negligible since $5^\circ$ precision in the antenna orientation is easily achievable.

For the effect of uncertainties on the antenna or amplifier gain $g(f)$, we consider two different effects: 
The gain may be off by a constant factor for all frequencies. In that case, the energy fluence of the reconstructed radio signal would change, but the frequency slope $m_f$ would remain unchanged. To test this, we scale the gain $g$ by a constant factor $p$:
\begin{equation}
    \label{eq:amp_scale}
    g^\prime=p\cdot g
\end{equation}

If the uncertainties in the amplifier gain are frequency-dependent, they may influence the reconstructed frequency slope $m_f$. To investigate this, the gain is modified by
\begin{equation}
    \label{eq:amp_slope}
    g^\prime(f)=g(f)\cdot\bigg(1+q\cdot\frac{f-\mathrm{200MHz}}{\mathrm{200MHz}}\bigg)
\end{equation}
Because we operate in the \SI{80}-\SI{300}{MHz} frequency band, this means the gain will be increased at one edge of the frequency band and decreased at the other by the same factor $q/2$. This modification will mainly affect the reconstruction of $m_f$, while the effect on $\Phi_E$ should be small.

The effect of these modifications to the amplifier gain are shown in Fig.~\ref{fig:hardware_errors}. On the left, the gain was scaled by $p=0.9$, on the right the slope was modified by $q=0.1$. In both cases, the median shifted by $\approx10\%$ compared Fig.~\ref{fig:reconstruction_quality}.

We tested changes in the amplifier gain for several values of p in the range $0.9<p<1.1$ and q in the range $-0.1<q<0.1$. In both cases, the shift of the median of the energy resolution is roughly linear with respect to p and q. The width of the Gaussian function fitted to the histogram showed little change, but the upper limit of the 68\% quantiles increased for some modified amplifier gains. This was caused by errors in determining if the air shower was seen from inside or outside the Cherenkov ring, which caused additional outliers. Under the assumption of a known shower geometry, the 68\% quantiles remained constant.

\section{Discussion}
We demonstrated that the signal detected by a single radio station is sufficient to accurately measure the energy in the electromagnetic part of an air shower. While the statistical uncertainty is better than 15\%, additional systematic uncertainties from the hardware response, namely their influence on the energy fluence and the frequency spectrum, have to be considered, underlining the importance of a thorough calibration.

The estimator was developed in the context of a surface station of a neutrino array, but the concept can be applied to different detectors, both for cosmic rays and neutrinos, as long as the signal is measured in a broad-band going beyond \SI{100}{MHz}.

\subsection{Implications for cosmic-ray detectors}
Our results may be applied to dedicated radio-based cosmic ray detectors in order to improve the accuracy of energy measurements. However, it requires a larger frequency band than the \SI{30}-\SI{80}{MHz} band that most radio detectors today operate in, for two reasons: A larger frequency band makes the measurement of the frequency slope parameter $m_f$ easier, as simply more data points are available for the reconstruction. Also, in the \SI{30}-\SI{80}{MHz} band, the frequency spectrum can not be described by a simple exponential function \cite{GrebePhD,JansenPhD}, so a more complex parameterization would have to be used, which makes the reconstruction more challenging. 
Together with an improved signal to noise ratio at higher frequencies \cite{SouthPoleGamma}, this makes the use of frequency bands outside the \SI{30}-\SI{80}{MHz} band an attractive option.

The method developed in this paper is especially well suited for sparse arrays (antenna spacing $\sim \mathcal{O}$(km)), where many events are only detected by a small number of stations. Sparse arrays also often focus on the detection of inclined air showers, for which this method yields the best results.

In addition, combining the radio measurement of a shower with the measurement of muons provides an excellent estimator for the primary particle. It is shown in \cite{Holt2019} that the ratio of the energy in the electromagnetic component of the shower, represented by the radio energy estimator, and the hadronic component, derived from the muonic component can used as powerful discriminator. Thus, building a hybrid detector of radio antennas and muon detector, may successfully address the composition problem. If using the energy reconstruction method presented here, the array can even be sparse and therefore low-cost and competitive. 
It is also worth noting that one of the major sources for uncertainties is confusion between showers seen from inside and outside the Cherenkov ring. For hybrid experiments combining radio with other detection methods, this issue may be trivial (see e.g.~\cite{TunkaRexSingleEnergy}). We also developed and parameterized this method for a detector located at the South Pole and at the site of the Pierre Auger Observatory in Argentina (see App.~\ref{sec:append_sites}).

\subsection{Implications for neutrino detectors}
The ability to measure the shower energy is a crucial requirement for any radio-based neutrino observatory. Experiments attempting to detect neutrino interactions in ice space their stations so far apart that a signal is expected to be only detected by a single station in most cases. This makes the ability to reconstruct the energy of a particle cascade from the measurement by one station alone essential.

By using air showers as a test case, we have explored a feasible way to achieve this. Because of the similarities between radio signals from air showers to those from neutrino-induced cascades in the ice, it is to be expected that our approach can be applied to neutrino detection as well. However, the radio signal from neutrinos shows additional complexities, such as an unknown vertex position, propagation effects through the ice and no dominant signal polarization due to a negligible geomagnetic emission. Therefore, further development is necessary in order to apply this method to neutrino detections. However, this method illustrates how a large band-width can be instrumental for the reconstruction and that methods like the forward folding of the pulse are robust to noise contributions. 

\section{Conclusions}
We have presented a method to reconstruct the energy in the electromagnetic cascade of an air shower using radio signals recorded in the $80-300$ MHz band by a single detector station. By using a forward folding technique, we are able to measure the frequency spectrum of the signal more accurately and use it to estimate the viewing angle in relation to the Cherenkov angle. This information is combined with the energy fluence and the signal direction to provide an energy estimator.

The accuracy of this method under realistic noise conditions and using a realistic triggering scheme was investigated. For events with a sufficiently high signal-to-noise ratio, statistical uncertainties around 10\% are achievable for the energy. Uncertainties in the orientation of the antennas were shown to be negligible while uncertainties in the amplifier gain have a significant effect and propagate linearly as systematic uncertainties on the reconstructed shower energy.

For cosmic-ray detectors, this method can be used to improve the accuracy of energy measurements, especially in sparse arrays, where often only measurements from a small number of stations are available. Regarding neutrino detectors, we have shown that it is possible to reconstruct the energy of a particle cascade from the signal of a single radio station, indicating a possible route towards the energy reconstruction for neutrinos at the highest energies.

\section{Acknowledgements}
We are thankful for the fruitful collaboration with our colleagues from the ARA and ARIANNA collaborations, especially for providing us with measured noise for this analysis. 
We acknowledge funding from the German research foundation DFG under grants NE 2031/2-1 and GL 914/1-1. 

\bibliographystyle{JHEP_2}
\bibliography{BIB}

\appendix

\section{Parameterization for other locations on Earth}
\label{sec:append_sites}

\begin{figure}
    \centering
    \includegraphics[width=\textwidth]{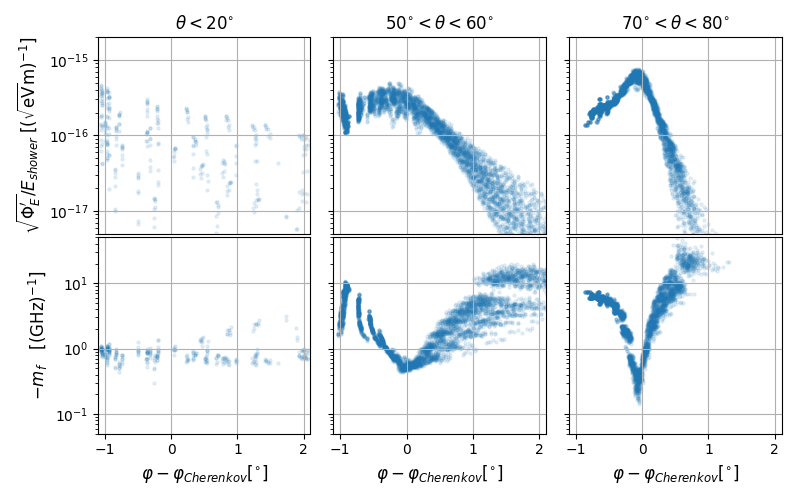}
    \caption{Lateral distributions of the energy fluence (top) and spectral slope (bottom) for simulated air showers at the South Pole at an altitude of \SI{2800}{m} above sea level. This figure is equivalent to Fig. \ref{fig:signal_ldf}, so a more detailed description can be found there.}
    \label{fig:ldf_southpole}
\end{figure}

Two additional tests were performed for other locations on Earth. The first location was the South Pole, which is the location of IceCube and its future extension IceCube-Gen2, which is planned to include a large radio array. Compared to Moore's Bay, the South Pole is at a higher elevation of \SI{2800}{m} above sea level. While the geomagnetic field is nearly vertical at both the South Pole and Moore's Bay, it is weaker at the South Pole. Fig.~\ref{fig:ldf_southpole} shows the dependence of the corrected energy fluence of the radio signal  $\Phi_E^\prime$ and spectral slope $m_f$ for different zenith angles for a radio detector at the South Pole. The distribution looks similar to the one shown in Fig.~\ref{fig:signal_ldf}, indicating that the same method will work. Because of the higher elevation, the shower maximum is much closer to the detector than at Moore's Bay and in some cases, the shower can even be clipped by the ground before reaching its maximum. Shower clipping is a known issue for the radio detection of air showers \cite{AERAPRD} and reduces the achievable accuracy, as an unknown fraction of energy has not been deposited in the shower. 

\begin{figure}
    \centering
    \includegraphics[width=\textwidth]{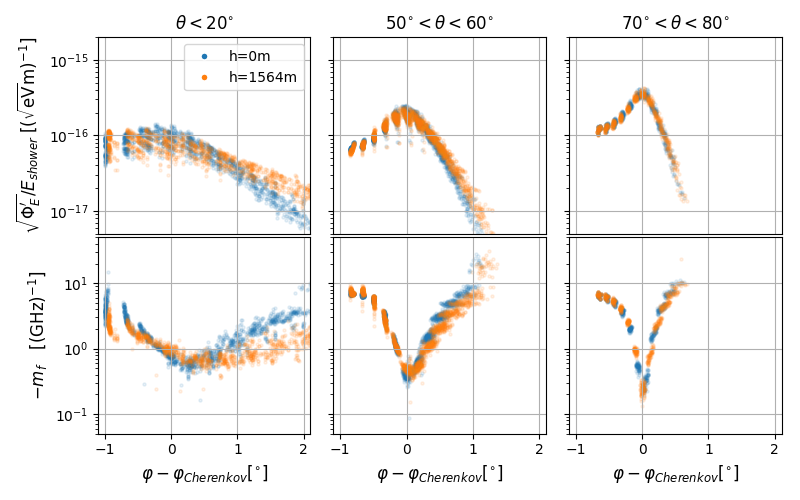}
    \caption{Lateral distributions of the energy fluence (top) and spectral slope (bottom) for simulated air showers at the site of the Pierre Auger Observatory. Simulations were done for detector stations at sea level (blue) and at an elevation of \SI{1564}{m} above sea level, the elevation of the Pierre Auger Observatory (orange). This figure is equivalent to Fig. \ref{fig:signal_ldf}, so a more detailed description can be found there.}.
    \label{fig:ldf_aera}
\end{figure}

The second location we simulated was the site of the Pierre Auger Observatory, where the Auger Engineering Radio Array (AERA) is located \cite{AERAPRD}. Also, the Auger Observatory is in the process of being upgraded with radio antennas at all surface detectors \cite{AugerPrime}. The two main differences compared to Moore's Bay are the higher elevation of roughly \SI{1500}{m} and the geomagnetic field, which has a zenith angle of \SI{54}{^\circ} and is weaker compared to Moore's Bay. Fig.~\ref{fig:ldf_aera} shows the dependence of the corrected energy fluence $\Phi_E^\prime$ and spectral slope $m_f$ on the viewing angle for a station at an elevation of \SI{1564}{m}, where the Pierre Auger Observatory is, and at sea level (same longitude and latitude). As one can see, the differences in elevation only have a minor influence. The overall trend is again the same, one even observes less scatter at small zenith angles.

For a station in Antarctica, the energy of an air shower with a small zenith angle is difficult to reconstruct, because two effects coincide: The steep and featureless lateral distribution at small zenith angles and an almost vertical magnetic field. For small zenith angles, small angles $\alpha$ to the geomagnetic field cause the radio signal to no longer be rotationally symmetric around the shower axis. In addition, the lateral distribution of $\Phi_E^\prime$ and $m_f$ is flatter and provides less discrimination power for the position with respect to the shower axis. Because of the inclined magnetic field at the Auger site, the scatter at small zenith angles is smaller. Small values of $\alpha$ instead occur for more inclined showers, where they pose a smaller problem because of the more pronounced dependence of $\Phi_E^\prime$ and $m_f$ on the viewing angle.

\begin{table}
    \centering
    \begin{tabular}{|c|c|c|c|c|c|}
        \hline
        & \multicolumn{3}{c|}{$A = (a\cdot \theta^2 + b \cdot \theta + c)\cdot\frac{10^{18}}{m\sqrt{eV}}$} & \multicolumn{2}{c|}{$s = m\cdot \theta + n$}\\
        & a & b & c & m & n \\
        \hline
         Moore's Bay & & & & & \\
         $\varphi < \varphi_{Cherenkov}$ & 442.46 & -281.75 & 324.96 & -0.1584 & -0.07943 \\
         $\varphi > \varphi_{Cherenkov}$ & 394.08 & -308.36 & 436.30 & 0.8070 & -1.4098 \\
         \hline
         South Pole & & & & & \\
         $\varphi < \varphi_{Cherenkov}$ & 976.30 & -1213.43 & 626.98 & -0.2273 & 0.05627 \\
         $\varphi > \varphi_{Cherenkov}$ & 643.39 & -667.08 & 478.06 & 1.3372 & -2.1653 \\
         \hline
         Auger & & & & & \\
         $\varphi < \varphi_{Cherenkov}$ & 229.96 & -123.75 & 110.51 & -0.1445 & -0.09820 \\
         $\varphi > \varphi_{Cherenkov}$ & 214.46 & -111.01 & 119.18 & 0.5936 & -1.1763 \\
         \hline
    \end{tabular}
    \caption{Zenith dependence of the parameters $A$ and $s$ from Eq.~(\ref{eq:energy_fluence_parametrization}) for a station at Moore's Bay, the South Pole and the site of the Pierre Auger Observatory.}
    \label{tab:parametrizations}
\end{table}

The parametrizations shown in Eq.~(\ref{eq:energy_fluence_parametrization}) and Fig.~\ref{fig:fit_parameters} have been calculated for the South Pole and the Auger site as well and are shown in Tab.~\ref{tab:parametrizations} along with the ones for Moore's Bay.

A detailed investigation of the achievable energy resolution at these locations would go beyond the scope of this paper, but it stands to reason that our method also works elsewhere and might even be able to provide better energy measurements for small zenith angles at locations with a less vertical magnetic field.

\section{Monte Carlo Data set}
\label{sec:append_MCset}
\begin{figure}
    \centering
    \includegraphics[width=\textwidth]{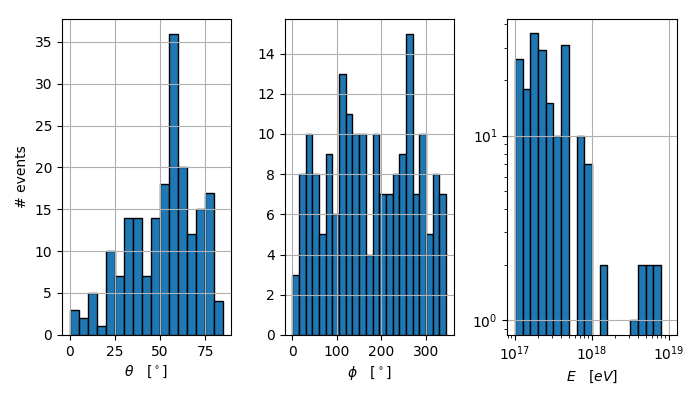}
    \caption{Zenith angles (left), azimuth angles (center), and energies (right) of the Monte Carlo data set used to evaluate the performance of the energy estimator.}
    \label{fig:event_properties}
\end{figure}

The Monte Carlo data set to evaluate the performance of the energy estimator was generated by randomly generating an incoming direction for the cosmic ray from an isotropic distribution and an energy from a distribution with a spectral index of $-2$. Out of a set of CoREAS files, the one with properties most similar to those generated was selected for each generated event. The resulting distribution of zenith angles, azimuths and energies is shown in Fig.~\ref{fig:event_properties}.

In each individual CoREAS file, observers are placed on a star-shape like pattern. Observers are placed symmetrically around the shower axis on the $\vec{v}\times\vec{B}-$ and the $\vec{v}\times(\vec{v}\times\vec{B})-$ axis, as well as on two lines bisecting the angle between the two axes.

\end{document}